\documentstyle[aps,epsf,twocolumn,prl]{revtex}
\newcommand{\avg}[1]{\langle{#1}\rangle}
\newcommand{\req}[1]{(\ref{#1})}
\newcommand{\beq}{\begin{equation}}
\newcommand{\eeq}{\end{equation}}
\newcommand{\beqar}{\begin{eqnarray}}
\newcommand{\eeqar}{\end{eqnarray}} 

\begin{document}
\title{Probability distribution of drawdowns in risky investments.}
\author{Sergei Maslov$^{(1,2)}$ and Yi-Cheng Zhang$^{(2)}$}
\address{$^1$Department of Physics, Brookhaven National
Laboratory, Upton, New York 11973, USA}
\address{$^2$Institut de Physique Th\'eorique,
Universit\'e de Fribourg P\'erolles, Fribourg CH-1700, 
Switzerland}
\date{\today}
\maketitle
\abstract{We study the risk criterion for investments 
based on the drawdown from the maximal value of the 
capital in the past. Depending on investor's
risk attitude, thus his risk exposure, we find that 
the distribution of these drawdowns follows a general 
power law. In particular, if the risk exposure is 
Kelly-optimal, the exponent of this power law 
has the borderline value of 2, i.e. the average 
drawdown is just about to diverge.}
\vskip 1cm

For repeated investments one may consider maintaining 
a fixed fraction $f$ of one's capital in risky assets, while 
keeping the rest of the capital in risk-free securities.
After the pioneering work by Kelly \cite{Kelly56} 
it is generally believed that the optimal strategy is to 
choose the investment fraction $f^*$ which 
maximizes the average growth rate of the logarithm of the capital
\cite{Breiman60,Thorp71,Maslov98}.
However, many economists and prudent 
investors would balk at this aggressive proposal, since
this strategy proposes a unique recipe for all purposes, without 
considering the investor's risk
profile. Traditional alternative to the Kelly 
investment recipe is to select the investment fraction,  
maximizing the expectation value of some investor-specific 
utility function \cite{Samuelson69}.
Unfortunately, this recipe leads to incorrect expectations, 
since for very broad distributions, such as 
log-normal distributions in multiplicative stochastic processes, the
expectation value is dominated by an exponentially small fraction of
outcomes, and is unlikely to be achieved after a reasonably large number 
of trials. In the past this common-sense observation has caused 
persistent debates and was often leading to fallacious conclusions 
\cite{Samuelson71}.
 
If the overall shape of utility function is of little relevance in
determining the optimal long-term investment strategy, what 
investment property one should consider to distinguish between aggressive and
conservative investment strategies? In this work we systematically
study a risk criterion based on the probability distribution
of drawdowns  of the capital measured relative to its highest 
value in the past, which we refer to as drawdowns from the maximum. 
It is often quoted in the trading 
community that the probability of a given 
drawdown from the maximum is one of the most sensible
parameters of an investment strategy \cite{Vince92}. 
Often investors identify their wealth as the highest achieved amount. 
Hence, at any time the current drawdown from the highest capital in the past 
gives a measure of investor's frustration, tests his strength of nerves and 
his faith in the ultimate recovery.

The definition of the drawdown from the maximum is rather natural.
Let $W(t)$ to denote investor's capital  
as a function of time. Define $W_{max}(t)$ to be  
the overall maximum of the capital up to this point in time:
$W_{max}(t)=\max _{t' \leq t} W(t')$. 
The current {\it drawdown from the maximum} 
(DDM) $D(t)$ is given by  
\beq
 D(t)=W_{max} (t)/W(t). 
\eeq
From this definition it follows that 
$D(t) \geq 1$ with equality realized only if the current capital 
is at its overall maximal value. 

Let us first find out the DDM probability distribution 
in a very general case where the investor's capital follows a 
discrete-time multiplicative random walk
\beq
W(t+1)=e^{\eta (t)} W(t) . 
\eeq 
In this expression a random number $\eta (t)$ is drawn 
at each time step $t$ from a given probability distribution 
$\pi(\eta)$. 
As usual, it is easier to work with the 
logarithm of the capital $h(t)=\ln W(t)$, which 
performs an ordinary random walk 
\beq
h(t+1)=h(t)+\eta (t) . 
\eeq 
The logarithmic drawdown from the maximum (LDDM) $LD(t)=\ln D(t)$ 
is simply given by 
\beq
LD(t)=\max _{t' \leq t} h(t') - h(t) . 
\eeq

To the purpose of calculating the probability distribution 
function of LDDM let us divide the time axis into a sequence of 
time intervals during which $h_{max}(t)=\ln W_{max}(t)$ 
stays constant (see Fig. 1). 
\begin{figure}
\epsfxsize=\columnwidth
\centerline{\epsfbox{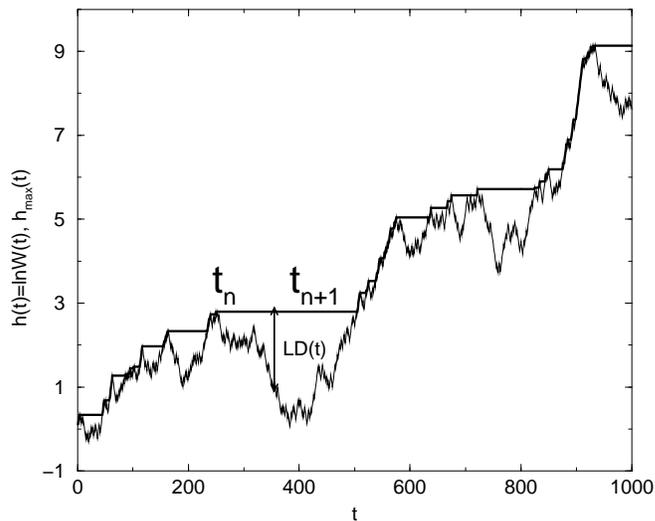}}
\caption{Random walk $h(t)=\ln W(t)$ (solid line) and its 
maximal value up to time $t$, $h_{max}(t)$ (bold solid line). Between
$t_n$ and $t_{n+1}$, the maximal value 
$h_{max}(t)$ stays constant and is equal to $h(t_n)$.}
\end{figure}
Each such interval starts at time $t_n$
when the walk is at the overall maximum of $h(t)$ and ends at time $t_{n+1}$ 
when this maximum is surpassed and replaced by a new, higher one. 
The motion in each of the intervals $(t_n,t_{n+1})$ can be viewed 
as a random walk with an {\it open } upper boundary at $h(t_n)$: 
the process ends (and the new one starts) 
when the walk leaves the interval $(-\infty, h_{max})$.
Within a single interval 
without loss of generality we can set $h_{max}=0$ 
and, consequently, $LD(t)=-h(t)$. 
In order to find the distribution of drawdowns from the 
maximum we need first to calculate the time dependent density 
$\rho( h,t)$ of the ensemble of random walks in the interval 
$(-\infty, 0)$ with an open upper boundary. The process starts 
at $t=0$, when $\rho(h,0)=\delta (h)$. 
Note that, since walks can leave the system,  
the probability of finding a walk within the interval $(-\infty, 0)$ 
(i.e. of finding the current maximum unsurpassed $t$ time steps after 
it was realized), $p_{\rm tot}(t)=\int_{-\infty}^{0}\rho (h,t)dh \leq 1$, 
is not conserved. Open boundary conditions are equivalent to maintaining 
$\rho(h,t)=0$ for $h \geq 0$ at all times.

The distribution of drawdowns from the 
maximum $P(x)dx ={\rm Prob}(x<LD<x+dx)$ measures the probability of finding 
a given logarithmic drawdown at an arbitrary time without any reference 
to the time $t$ elapsed since the last maximum.
Therefore, $P(x)$ is proportional to the time-cumulative 
density at the point 
$h_{max}-x=-x$: $P(x) \sim \sum _{t=0}^{\infty} \rho (-x,t)$.  
Including normalization one gets
\beq
P(x) ={\sum _{t=0}^{\infty} \rho (-x,t) \over 
\int _0^{\infty} \sum _{t=0}^{\infty} \rho (-x',t) d x'} . 
\eeq

Any ensemble density $\rho (x,t)$  should 
obey the following integral equation of motion 
\beq
\rho (x,t)=\int \rho (x-\eta, t-1) \pi(\eta) d \eta
\label{e.motion}
\eeq
This equation expresses the density $\rho(x,t)$
at time $t$ in terms of the known density at the previous time step
$t-1$ and the probability distribution of jumps. It fixes $\rho (x,t)$  
for $x \leq 0$, while the definition of 
an open boundary random walk requires $\rho (x,t)=0$ for $x \geq 0$. 
The stationary probability distribution of 
drawdowns from the maximum makes sense only for random walks 
with a positive (upwards) drift $\avg{\eta}>0$. Indeed, for a negative
(downwards) drift the maximum realized in the beginning of the 
process is likely to be never surpassed.  
In this case as the walk drifts further and further down it 
samples larger and larger drawdowns so that the probability 
distribution of drawdowns never becomes stationary.  
Using the equation of motion \req{e.motion} 
for $\rho (h,t)$ one gets $P(x)=A^{-1} \sum _{t=0}^{\infty} \rho (-x,t)=
A^{-1} (\sum _{t=1}^{\infty} \int \rho (-x-\eta,t-1) \pi(\eta) d \eta 
+ \rho (-x,0))= \int P(x+\eta) \pi(\eta) d \eta + A^{-1} \delta (x)$. 
Here $A= \sum _{t=0}^{\infty} \int _0^{\infty} \rho (-x',t) d x'$
is the normalization factor.

Therefore, for  $x>0$ one has
\beq
P(x)=\int _{-x}^{\infty} P(x+\eta) \pi(\eta) d \eta . 
\label{e.stationary}
\eeq 
Note also 
that in order to follow our definition of logarithmic drawdowns 
we had to change the sign in front of $\eta$, compared to that in 
Eq. \ref{e.motion}. It is a straightforward task to determine 
the asymptotic behavior of $P(x)$ for 
$x$ much bigger than the typical value of $\eta$. 
In this case we can safely disregard that 
$P(x)=0$ for $x<0$ and plug the ansatz functional form 
$P(x) \sim \exp (-\Gamma x)$ into the Eq. \ref{e.stationary}. 
The parameter $\Gamma>0$ is then determined from the equation
\beq
\int_{-\infty}^{\infty} \pi(\eta) \exp (-\Gamma \eta) d \eta =1 . 
\label{e.gamma}
\eeq
One can show that this equation has at most one strictly positive 
solution. In fact the sufficient and necessary condition 
for the existence of such solution is a 
positive upwards drift $\avg{\eta}>0$ plus a  
nonzero support of $\pi(\eta)$ for negative $\eta<0$.
Indeed, first one notices that the second derivative 
of $V(\Gamma)$, where 
$V(\Gamma)$ is the LHS of Eq. \ref{e.gamma}   
with respect to $\Gamma$, is strictly positive, and $\Gamma=0$ is 
the obvious solution to $V(\Gamma)=1$. 
The nonzero $\pi(\eta)$ for some $\eta<0$ 
guarantees that $V(+\infty)=+\infty$.  Since 
$d V/d \Gamma |_{\Gamma=0}= -\avg{\eta}<0$, the 
continuity of $V(\Gamma)$ guarantees the existence of 
the positive solution of $V(\Gamma)=1$. The positive 
second derivative ensures its uniqueness.

In order to get a better feeling of how the parameters of the jump 
distribution $\pi(\eta)$ determine $\Gamma$ we consider 
two particular functional forms of the distribution 
$\pi(\eta)$. We first see what happens if $\pi(\eta)$ 
has a binomial shape.
For simplicity let us take a particular binomial distribution, 
where $\eta=\ln \Lambda$ with probability $p>1/2$, and 
$\eta=-\ln \Lambda$ with probability $1-p$. 
In other words, with probability $p$ one's capital 
is multiplied by $\Lambda>1$, otherwise it is divided by
$\Lambda$. For this distribution 
Eq. \ref{e.gamma} reduces to 
$p/y+(1-p)y=1$, where $y=\Lambda^{\Gamma}$.
This quadratic equation has two solutions $y_1=1$ 
and $y_2=p/(1-p)$. For $p>1/2$ (upwards drift condition) 
the second solution gives the desired positive 
\beq 
\Gamma _{\rm binomial}={\ln p - \ln (1-p) \over \ln \Lambda}.
\eeq

The other case we use to illustrate Eq. \ref{e.gamma} 
is when $\pi(\eta)$ has a Gaussian shape 
$\pi(\eta)=(1/\sqrt{2 \pi} \sigma) \exp (-(\eta-\mu)^2/2\sigma^2)$.  
Then Eq. \ref{e.gamma} can be rewritten as 
$\exp( -\Gamma(\mu-\Gamma \sigma^2/2))=1$. The unique 
nontrivial ($\Gamma \neq 0$) solution, given by
\beq
\Gamma _{\rm gaussian}={2\mu \over \sigma^2} 
\label{e.gamma_gauss}
\eeq 
is positive, provided $\mu>0$, i.e. the random walk 
has an upwards drift.

The last equation can be also derived within a continuous time approach. 
Indeed, increments of a continuous-time random walk, taken at a discrete 
time intervals, necessarily have a gaussian shape so that Eq. 
\ref{e.gamma_gauss} should hold in this case. 
Another way to see this is to replace the integral equation 
\req{e.stationary} with the differential stationary 
Fokker-Planck equation  
\beq
\mu {\partial P(x) \over \partial x}+
{\sigma^2 \over 2} {\partial^2 P(x) \over \partial x^2}=0, 
\eeq
where $\mu$, and $\sigma ^2$ 
are the drift  velocity and the dispersion of the random walk.
This equation has a solution $P(x)=A^{-1} \exp(-\Gamma x)$, where 
$\Gamma$ is given by
\beq
\Gamma_{\rm continuous}={2\mu \over \sigma^2} .
\label{e.gamma_cont}
\eeq 

The exponential distribution 
of logarithmic drawdowns $LD=\ln D$ corresponds to the power law
distribution of  drawdowns themselves. To properly change 
variables one notices that ${\rm Prob}(LD>x) \sim \exp (-\Gamma x)$.
Therefore, ${\rm Prob}(D>y) \sim y^{-\Gamma}$, and for the 
distribution of $D$ one has $P(D) \sim 1/D^{1+\Gamma}$. 

It is interesting to note that the mechanism by which drawdowns 
from the maximum acquire a power law distribution is similar to 
that of the multiplicative random walk pushed against the wall. 
This mechanism, which was first analyzed in a 
financial context by Levy \& Solomon \cite{Solomon96} 
and later studied in greater detail in \cite{Sornette97,Marsili98}, 
is rather simple. It is well known that the problem of finding a stationary
distribution of a multiplicative random walk drifting 
in the direction of a {\it reflecting} wall can be rewritten in terms 
of a Fokker-Planck equation  for the logarithm of the observable variable 
with reflecting boundary condition at the position of the wall.
The solution of this equation 
has the well known exponential (Boltzmann) form which, 
being rewritten in terms of the variable, subject to the
multiplicative noise, becomes a power law. In our analytical 
approach to the problem of drawdowns from the maximum, the current 
maximal value of the capital serves as an {\it absorbing} upper wall 
for a random walk (once the walk surpasses the current 
maximum, the value of the maximum has to be updated, which can be
looked at as simply taking another representative of the ensemble).
Of course, the Fokker-Plank
equation with an absorbing boundary, unlike with a 
reflecting boundary, does not allow
for a stationary solution. However, as was demonstrated above, 
the equations for cumulative (integrated over time) distributions 
are identical in both cases. That is why it should not be surprising 
that Eq. \ref{e.gamma_cont} of this paper is identical to the 
Eq. 10 of Ref. \cite{Marsili98}, 
which determines the exponent of the stationary power law distribution
of multiplicative random walk pushed against the hard wall.

Now we are in a position to derive the distribution 
of the drawdowns from the maximum for the investor, 
following a  
constant investment fraction strategy 
\cite{Kelly56,Breiman60,Thorp71,Maslov98}. 
In such a strategy the investor invests a fraction of 
his capital in one risky asset while keeping the remainder safely 
in risk-free securities.  At each discrete time step 
the investor sells or buys the correct amount of shares of 
risky asset to adjust the current value of his asset holdings to 
{\it precisely} the fraction $f$ of his total capital. 
This investment fraction $f$ (leverage factor if $f>1$) 
is the sole parameter defining the strategy. In this work we 
do not allow the change of the reinvestment time interval 
(the discrete time step at which the investor adjusts his asset 
holdings). Also for simplicity  we set the risk-free 
interest rate to zero. The generalization 
to a more general situation is rather straightforward, but 
makes our final formulas less transparent.
The evolution of investor's capital for a fixed investment 
fraction strategy is given by a multiplicative random walk 
\beq
W(t+1)=W(t) \left( 1-f+f e^{\eta(t)} \right) . 
\eeq
In this expression the random variable $\eta(t)$ describes the
multiplicative fluctuations 
of the price $p(t)$ of the risky asset: $p(t+1)=e^{\eta(t)} p(t)$. 
The results for the distribution of drawdowns from the maximum derived 
above apply to the fixed investment fraction strategy if one uses 
$f$-dependent random walk variable 
$\eta_f=\ln(1-f+f e^{\eta})$ so that 
$e^{\eta_f}=1-f+f e^{\eta(t)}$. If $f=1$, i.e. the whole 
capital is invested in risky asset, $\eta_f=\eta$ and 
the whole capital just follows the multiplicative random walk 
of the risky asset's price. It is clear that 
by selecting a smaller investment fraction $f$ the investor reduces 
the probability of significant drawdowns from the maximum, so that 
$\Gamma_f$ is a decreasing function of $f$.

The results are especially 
straightforward in the case when the logarithm of the stock price follows a 
continuous gaussian random walk with drift velocity 
$\mu$ and dispersion $\sigma^2$.
As it was shown for instance in \cite{Maslov98}, for a gaussian $\pi(\eta)$
the logarithm of the capital subject to a  
fixed investment faction strategy has the drift velocity  
$\mu _f$ and dispersion $\sigma _f^2$ given by
\beqar
\mu _f&=&\left( \mu+{\sigma^2 \over 2} \right) f-
{\sigma^2 f^2 \over 2} ; 
\label{e.mu_f} \\
\sigma  _f^2&=&\sigma^2 f^2
\eeqar
The exponent of the power law distribution of drawdowns $P(D)$,
$\tau _f=\Gamma _f+1=2 \mu _f/\sigma _f^2+1$, 
in this case is given by
\beq
\tau _f^{\rm gaussian}={2\mu + \sigma^2 \over \sigma^2 f}
\eeq
The bigger is this exponent, the safer is your investment 
from large drawdowns. Of course, the stationary distribution of 
drawdowns is limited to the case when $\mu _f>0$. For the exponent 
$\tau  _f$ this corresponds to the condition $\tau _f>1$, i.e. 
normalizable $P(D)$.

It was suggested by Kelly \cite{Kelly56} that for the long term 
investment the optimal fixed investment fraction strategy would be 
the one maximizing $\mu _f$.
For the Gaussian $\pi(\eta)$ the investment fraction $f^*$ 
in this Kelly-optimal strategy 
is given by $f^*=1/2+\mu/\sigma^2$. 
Indeed, this is what one gets from maximization 
of the drift velocity $\mu_f$ given by Eq. \ref{e.mu_f}.
It is interesting to note that
for the Kelly optimal strategy the DDM distribution exponent 
has a superuniversal value
\beq
\tau _{f^*}=2 .
\eeq 
This result is not restricted to gaussian $\pi(\eta)$.
It is straightforward to demonstrate that it holds at 
the Kelly optimum for any  $\pi(\eta)$.  
Indeed, by definition of the Kelly optimal strategy it maximizes
the growth rate of the logarithm of the capital given by 
$\mu _f=\avg{\ln(1-f+f e^\eta)}$. 
Therefore, $0=\partial \mu _f /\partial f |_{f^*}=
\avg{(e^\eta-1)/(1-f^*+f^* e^\eta)}=(1-\avg{(1-f^*+f^* e^\eta)^{-1}})/f^*$. 
From this equation it follows that at the Kelly optimum one has 
$\avg{e^{-\eta_f}}=\avg{e^{-\ln (1-f^*+f^* e^\eta)}}=
\avg{(1-f^*+f^* e^\eta)^{-1}}=1$, i.e. $\Gamma_{f^*}=1$ 
($\tau_{f^*}=\Gamma_{f^*}+1=2$) is the solution to \req{e.gamma}.
That proves 
that for an arbitrary distribution $\pi(\eta)$ 
 precisely at the Kelly optimum $f^*$  
the power law distribution of drawdowns has a superuniversal exponent
$\tau_{f^*}=2$.

Let us illustrate 
these results using an example of a risky asset, the price 
of which with equal probability $p=1/2$ 
goes up by 30\% 
or down by -24.4\%.
This is precisely the example of a hypothetical 
``red chip'' stock that we used in \cite{Maslov98} to illustrate
the power of Kelly optimization. The stock itself is doomed: 
its price is going down by roughly 1\% every time step 
(typically at each time step the price is multiplied 
by $\sqrt{1.30 \cdot 0.756} \approx 0.99$). On the other hand, since average 
return of $2.8 \%$ of this stock is positive, following the 
Kelly-optimal fixed investment 
fraction strategy with $f^* \simeq 0.3825$ results 
in a positive growth rate of investor's capital of some $0.53 \%$. 
We have simulated the outcomes of investment 
process with different investment fractions both above and 
below Kelly optimal. Fig.2 displays the time dependence 
of investor's capital for $f=0.1$, $0.38$, 
$0.7$, and $1$. 
\begin{figure}
\epsfxsize=\columnwidth
\centerline{\epsfbox{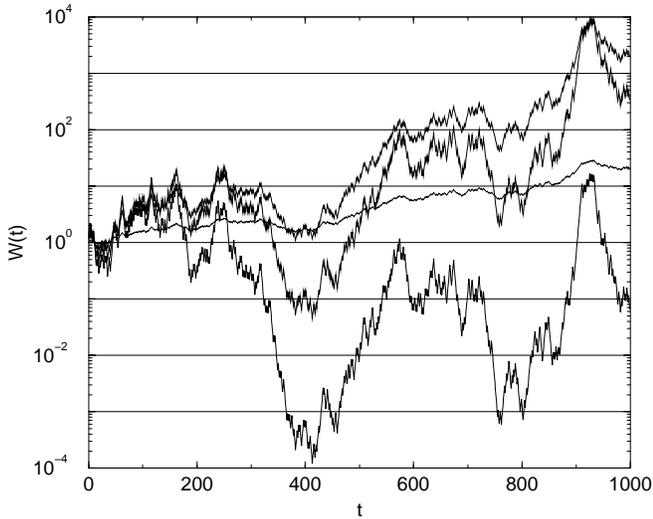}}
\caption{The evolution of the capital invested in 
the hypothetical risky asset described in the text 
at different investment fractions $f=0.38, 0.7, 0.1, 1$ 
(from top to bottom by the last point
$W(1000)$ in the time series) .}
\end{figure}
%
It is clear from this figure that 
the final capital after 1000 time steps grows as $f$ is increased 
from $0$ to $0.38$ and starts to go down above $0.38$ 
so that for $f>0.765$ the typical growth rate becomes negative and 
the investor ends up loosing money. This is illustrated in Fig. 2 
on the example of the $f=1$ curve, where the investor trusted his 
whole capital to the stock and is going down together with this 
doomed stock.

In Fig. 3 we plot the probability distributions 
of the drawdowns from the maximum for different 
investment fractions in this stock. 
\begin{figure}[t]
\epsfxsize=\columnwidth
\centerline{\epsfbox{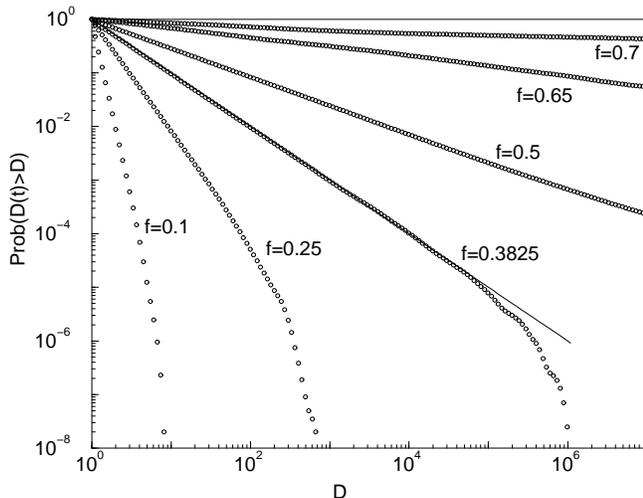}}
\caption{The probability to have a drawdown bigger than $D$ as a
function of $D$ for the same hypothetical risky asset as in 
Fig. 2. The power law exponent $\tau_f$ is systematically 
decreasing with the investment fraction $f$ ranging from $0.1$ 
to $0.7$. The exponent of the $P(D(t)>D)$ at the Kelly optimum 
$f^*=0.3825$ is in excellent agreement with the theoretical prediction
$\Gamma _{f^*}=1$. $5 \times 10^8$ data points were used to make
histograms  in this plot.}
\end{figure}
The trend of increasing probability of large drawdowns 
as $f$ is increased can be clearly seen. The P(D) 
calculated at the Kelly optimal fraction $f^* \simeq 0.3825$ is in 
agreement with our theoretical prediction of $\Gamma _{f^*}=1$.

To illustrate our results 
on a more concrete example we analyzed the time dependence of 
the capital invested in S\&P500 index during the year of 1996, 
using half hourly data provided by Olsen\&Associates.
In our hypothetical ``investment'' we selected and maintained 
on half-hour basis three different fixed leverage factors: 
$f=5$, 10, and 15. 
Any $f>1$, of course, can be realized only if such a 
leverage ratio is allowed (this is the case e.g. using
derivatives such as futures).
The resulting drawdown distributions are shown in Fig. 4.
\begin{figure}
\epsfxsize=\columnwidth 
\centerline{\epsfbox{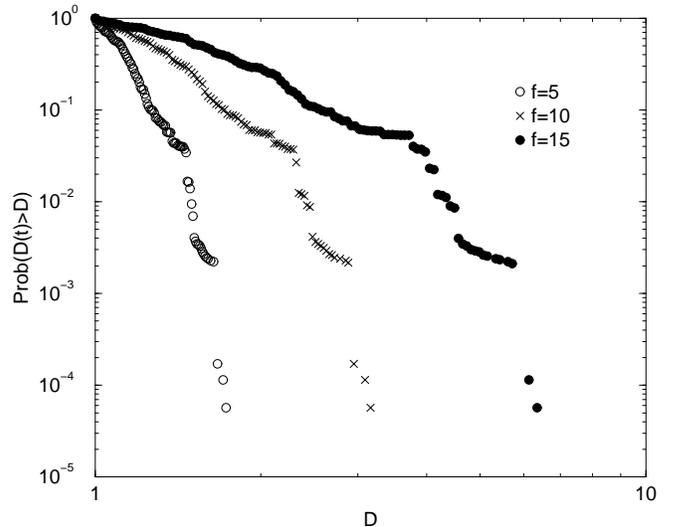}}
\caption{The probability to have a drawdown bigger than $D$ as a
function of $D$ for a leveraged investment in the 
S\&P500 index during the year of 1996. The investment (leverage) factors
are $f=5,10,15$ from left to right.}
\end{figure}
The largest leverage factor  $f=15$ approximately corresponds to 
the Kelly optimum for this asset 
under the condition of zero risk-free interest rate.

In summary, we have proposed and studied a risk measure for 
repeated investment games. We see that, unlike the traditional
expected utility approach describing the risk in 
terms of a single number, we need a whole function
to judge if the risk is worth undertaking. Under general 
conditions this function -- the 
distribution of drawdowns from the maximum -- 
has a power law shape. Kelly's optimal solution represents
the most aggressive strategy, since the power law barely gives a finite
expectation value of drawdowns (the exponent being 2). More 
risk-adverse investors can resort to sub-optimal strategies in the
Kelly sense, where large drawdowns
are considerably tamed. However, even those ``safer'' strategies are not
absolutely free from the risk: since power laws do not
have built in cutoffs, ruins (large drawdowns) 
can in principle arrive but are much less likely.

The work at Brookhaven National Laboratory was 
supported by the U.S. Department of Energy Division
of Material Science, under contract DE-AC02-98CH10886.
One of us (S.M.) thanks the Institut de Physique Th\'eorique,
Universit\'e de Fribourg, for the hospitality and financial support 
during the visit, when this work was performed.

\end{document}